\documentclass[12pt]{book}
\usepackage{plenum}
\begin{document}

\chapter{CORRELATIONS OF WAVE FUNCTIONS IN DISORDERED SYSTEMS
\protect\footnote{Talk given at the NATO Advanced Study Institute
``Supersymmetry and Trace Formulae'', Isaac Newton Institute,
Cambridge, UK, 8--19 September 1997.}}

\author{Alexander D. Mirlin \protect\footnote{Also at Petersburg
Nuclear Physics Institute, 188350 St. Petersburg, Russia.}}
\affiliation{Institut f\"ur Theorie der kondensierten Materie,
Universit\"at Karlsruhe, 76128 Karlsruhe, Germany}

\section{INTRODUCTION}

Statistical properties of eigenfunction amplitudes in disordered and
chaotic systems have attracted considerable research interest
recently. Fluctuations of the wave function amplitudes are believed to
determine statistical properties of conductance peaks in quantum
dots in the Coulomb blockade regime \refnote{\cite{Jal,PEI,MPA,Chang,Folk}},
and can be directly measured in the microwave cavity experiments
\refnote{\cite{Stockmann,Sridhar}}. On the theoretical side, the
recent progress 
is based on application of the supersymmetry method to the problem of
eigenfunction statistics \refnote{\cite{MF1,PEI}}. It was found
that in so-called zero-mode approximation
the distribution of eigenfunction amplitudes 
is correctly described by formulas of the
random matrix theory (RMT). Deviations from the RMT predictions were
studied in Refs.\cite{MF1,MF,tails,ADM-tails}.

The present article addresses a problem of correlations of
eigenfunction amplitudes. A description of correlations in amplitudes
of a wave function in relatively close points of a chaotic billiard
was proposed by Berry \refnote{\cite{Berry}} within a RMT-like 
assumption that the wave
function is a superposition of plane waves with random coefficients.
More recently, these correlations were considered in a disordered
system within the zero-mode approximation \refnote{\cite{Prig}}; the
result  
was later shown \refnote{\cite{Srednicki}} to be equivalent to that of
Ref.\cite{Berry}. This result is valid for small separations of the
two points (less than the mean free path $l$). Here we consider such
correlations for arbitrary distances. These correlations determine, in
particular, fluctuations of matrix elements of the (Coulomb)
interaction, which are in turn important for statistical properties of
spectra of quantum dots. Another topic addressed in the present
paper is that of correlation of amplitudes of different
eigenfunctions. Such correlations, while absent in RMT, appear in a
disordered system. They become especially strong near the 
Anderson metal-insulator transition, where they play a crucial role in
supporting the RMT-like level repulsion.

This article is based on recent works done in collaboration
with Ya.~M.~Blanter, Y.~V.~Fyodorov, and B.~A.~Muzykantskii
\refnote{\cite{MF,BM,BM1,FM2,BMM,BMM2}}. 

\section{EIGENFUNCTION CORRELATIONS IN THE WEAK LOCALIZATION REGIME}

In this section, we study the correlations of eigenfunctions in the
regime of a good conductor \refnote{\cite{MF,BM,BM1}}.
The correlation function of amplitudes of one and the same
eigenfunction with energy $E$ can be formally defined as follows:
\begin{equation}
\label{e1}
\alpha({\bf  r}_1, {\bf  r}_2, E) = \left\langle
\vert \psi_k({\bf  r}_1) \psi_k({\bf  r}_2) \vert^2
\right\rangle_{E}\equiv
\Delta \left\langle \sum_k \vert \psi_k({\bf 
r}_1) \psi_k({\bf  
r}_2) \vert^2 \delta (E- \epsilon_k) \right\rangle,
\end{equation}
where $\psi_k ({\bf r})$ and $\epsilon_k$ are eigenfunctions and
eigenvalues of the Hamiltonian in a particular
disorder configuration $U({\bf r})$, the angular brackets
 $\langle\ldots\rangle$
denote averaging over the disorder potential $U({\bf  r})$, and 
$\Delta=\langle \sum_k \delta (E - \epsilon_k) \rangle^{-1}$ is
the mean level spacing. In the case of a 
real or numerical experiment, calculation
of the correlation function (\ref{e1}) would include averaging over
all states in a relatively narrow energy window around $E$.
An analogous correlation function for two
different eigenfunctions is defined as
\begin{eqnarray}
\label{e2}
\sigma({\bf  r}_1, {\bf  r}_2, E, \omega) &=&
\left\langle \vert \psi_k({\bf  r}_1) \psi_l ({\bf  r}_2)
\vert^2 \right\rangle_{E,\omega} \nonumber\\
&\equiv & \Delta^2 R_2^{-1}(\omega)
\left\langle \sum_{k\ne l} \vert \psi_k({\bf 
r}_1) \psi_l({\bf   
r}_2) \vert^2 \delta (E - \epsilon_k) \delta(E + \omega
- \epsilon_l) \right\rangle,
\end{eqnarray}
where $R_2(\omega)$ denotes the two-level correlation function,
\begin{equation}
\label{e3}
R_2(\omega) = \Delta^2 \left\langle \sum_{k,l} \delta(E -
\epsilon_k) \delta(E + \omega - \epsilon_l) \right\rangle.
\end{equation}
Eq.(\ref{e2}) defines an overlap of the eigenfunctions
$\psi_k$ and $\psi_l$ provided they have energies  close to $E$ 
with the energy difference equal to $\omega$.

To evaluate $\alpha({\bf  r}_1, {\bf  r}_2, E)$ and
$\sigma({\bf  r}_1, {\bf  r}_2, E, \omega)$ (Ref.\cite{BM1}), we employ an
identity    
\begin{eqnarray}
&& 2\pi^2\left[\Delta^{-1}\alpha({\bf r}_1, {\bf r}_2, E)
\delta(\omega)+\Delta^{-2}
\tilde{R}_2(\omega)\sigma({\bf  r}_1, {\bf  r}_2, E,\omega)\right]
 \label{e4}\\
&&= \mbox{Re}\left[\left\langle
G^R({\bf  r}_1, {\bf  r}_1, E)G^A({\bf  r}_2, {\bf  r}_2, E+\omega)  -
 G^R({\bf  r}_1, {\bf  r}_1, E)
G^R({\bf  r}_2, {\bf  r}_2, E+\omega)\right\rangle \right] \nonumber,
\end{eqnarray}
where $G^{R,A}({\bf  r}, {\bf  r'}, E)$ are retarded and advanced
Green's functions and 
$\tilde{R}_2(\omega)$ is non-singular part of the level-level
correlation function:
$R_2(\omega)=\tilde{R}_2(\omega)+\delta(\omega/\Delta)$.
A natural question, which arises at this point, is whether the
r.h.s. of Eq.(\ref{e4}) cannot be simply found within the
diffuson-Cooperon perturbation theory \refnote{\cite{AShk}}. Such a
calculation would, however, be justified only for $\omega\gg\Delta$
(more precisely, one has to introduce an imaginary part of frequency:
$\omega\to\omega +i\Gamma$, and require that
$\Gamma\gg\Delta$). Therefore, it would only allow to find 
a smooth in $\omega$ part of
$\sigma({\bf  r}_1, {\bf  r}_2, E, \omega)$ for $\omega\gg \Delta$. 
Evaluation of $\alpha({\bf  r}_1, {\bf  r}_2, E)$, as well as of
$\sigma({\bf  r}_1, {\bf  r}_2, E, \omega)$ at $\omega\sim\Delta$
cannot be done within such a calculation. For this reason, we employ a
non-perturbative supersymmetry approach below.

The r.h.s. of Eq.(\ref{e4}) can be expressed in terms of the
supermatrix $\sigma$-model \refnote{\cite{Efetov}}, yielding:
\begin{eqnarray} \label{e5}
& & 2\pi^2 \left[\Delta^{-1} {\alpha({\bf r}_1, {\bf r}_2,
E)} \delta(\omega) + \Delta^{-2}
{\sigma({\bf r}_1, {\bf r}_2,E, \omega)}  
\tilde{R}_2(\omega) \right] \nonumber \\
& & = (\pi\nu)^2 \left[1-\mbox{Re} \langle  Q_{bb}^{11}
({\bf  r}_1) Q_{bb}^{22} ({\bf  r}_2) \rangle_F 
 - k_d({\bf r}_1-{\bf r}_2) \mbox{Re} \langle Q^{12}_{bb}
({\bf  r}_1) Q^{21}_{bb}({\bf  r}_1) \rangle_F \right],
\end{eqnarray}
where $k_d({\bf r}) = (\pi\nu)^{-2} \langle\mbox{Im}G^R
({\bf r})\rangle^2$ is a short-range function  explicitly given by 
\begin{equation}
 k_d({\bf r}) =  \exp (-r/l) \left\{
\begin{array}{ll} 
J_0^2(p_F r), & \ \ 2D\\
(p_Fr)^{-2} \sin^2 p_Fr, & \ \ 3D 
\end{array}
\right..
\label{e6}
\end{equation}
Here $\langle \dots \rangle_F$ denotes the averaging with the action
of the supermatrix sigma-model $F[Q]$:
\begin{eqnarray} \label{e7}
& &\langle \dots \rangle_F = \int DQ ( \dots ) \exp(-F[Q]), \nonumber \\ 
& & F[Q] = - \frac{\pi\nu}{4} \int d{\bf r} \ \mbox{Str} [D(\nabla
Q)^2 + 2i(\omega + i0) \Lambda Q],
\end{eqnarray}
where $D$ is the diffusion coefficient, $\nu$ is the density of states,
$Q = T^{-1}\Lambda T$ is a 4$\times$4 supermatrix, $\Lambda =
\mbox{diag} (1,1,-1,-1)$, and $T$ belongs to the supercoset space
$U(1,1 \vert 2)/U(1\vert 1)\times U(1 \vert 1)$. The symbol
$\mbox{Str}$ denotes the supertrace defined as 
$\mbox{Str} B = B_{bb}^{11} - B_{ff}^{11} + B_{bb}^{22} - B_{ff}^{22}$. The 
upper matrix indices correspond to the retarded-advanced
decomposition, while the lower indices denote the boson-fermion
one. The action (\ref{e7}) is written for the case of so-called
unitary ensemble (broken time reversal symmetry), which we consider
below. Generalization to a system with time reversal symmetry
(orthogonal ensemble) is straightforward, and the results are
presented in the end of the section. 
Evaluating the $\sigma$-model correlation functions in the
r.h.s. of Eq.(\ref{e5}) and separating the result into the singular 
the singular (proportional to $\delta(\omega)$) 
and regular at $\omega = 0$ parts, one can obtain the correlation functions
$\alpha({\bf r}_1,{\bf r}_2,E)$ and
$\sigma({\bf r}_1,{\bf r}_2,E,\omega)$.
The two-level correlation function $R_2(\omega)$ entering
Eq.(\ref{e5}) is given in this formalism by
\begin{equation}
R_2(\omega)={1\over 2V^2}\mbox{Re}\int d{\bf r_1}d{\bf r_2}
\left[1-\langle Q_{bb}^{11}({\bf r}_1) Q_{bb}^{22} ({\bf  r}_2)
\rangle_F \right].
\label{e7a}
\end{equation}

In the metallic (weak localization) regime, 
the sigma-model correlation functions $\langle Q_{bb}^{11}({\bf 
r}_1) Q_{bb}^{22} ({\bf  r}_2) \rangle_F$ and $\langle
Q_{bb}^{12}({\bf  r}_1) Q_{bb}^{21} ({\bf  r}_2) \rangle_F$ can be
calculated for relatively low frequencies $\omega \ll E_c$ with the
use of a general method developed in Refs.\cite{KM,MF}, which allows
one to take into account spatial variations of the field $Q$. The
results are obtained in form of expansions in $g^{-1}$, where $g$ is
the dimensionless conductance. First, we
restrict ourselves to the terms of order $g^{-1}$. Then, the result
for the first correlation function reads as  
\begin{equation} \label{1122a}
 \langle Q_{bb}^{11}({\bf  r}_1) Q_{bb}^{22} ({\bf  r}_2)
\rangle_F 
= -1 - 2i {\exp (i\pi s) \sin \pi s\over (\pi s)^{2}} - 
\frac{2i}{\pi s}
\Pi({\bf  r}_1, {\bf  r}_2)\ ,
\end{equation}
where $s=\omega/\Delta+i0$.
Here the diffusion propagator $\Pi$ is the solution to the diffusion
equation
\begin{equation} \label{diff} 
- D \nabla^2\Pi({\bf  r}_1,{\bf  r}_2) = (\pi \nu)^{-1}
[\delta({\bf  r}_1 - {\bf  r}_2)-V^{-1}]   
\end{equation}
with the Neumann boundary condition (normal derivative equal to
zero at the sample boundary), which can be presented in the form
\begin{eqnarray} \label{pi1}
\Pi({\bf  r}_1,{\bf  r}_2) & = & (\pi \nu )^{-1}
\sum_{{\bf q}} (Dq^2)^{-1} \phi_{{\bf q}} ({\bf  r}_1)
\phi_{{\bf q}} ({\bf  r}_2), 
\end{eqnarray}
with $\phi_{{\bf q}}$ being the eigenfunction of the
diffusion operator corresponding to the eigenvalue $Dq^2$, 
${\bf q} \ne 0$. 
The first two terms in Eq. (\ref{1122a}) represent the result of
the zero-mode approximation \refnote{\cite{Efetov}}, which takes into
account only the spatially constant configurations of the
field $Q({\bf  r})$, so that the
functional integral over $DQ({\bf  r})$ is reduced to an integral
over a single matrix $Q$. The last term is the correction of order
$g^{-1}$. An analogous calculation for the second correlator yields:
\begin{equation} \label{Q3}
\langle Q_{bb}^{12} ({\bf  r}_1) Q_{bb}^{21} ({\bf  r}_2)
\rangle_F =  -2\left\{ {i\over \pi s }
+ \left[ 1 +
 i { \exp (i\pi s) \sin \pi s\over (\pi s)^{2}} 
\right] \Pi({\bf  r}_1, {\bf  r}_2) \right\}.
\end{equation}
Now, separating regular and singular parts in r.h.s. of Eq. (\ref{e5}),
we obtain the following result for the autocorrelations of the same 
eigenfunction:
\begin{equation} \label{fin1}
 V^2\langle \vert \psi_k({\bf  r}_1) \psi_k({\bf  r}_2) \vert^2
\rangle_{E} -1  
 =   k_d(r) 
 [1 + \Pi({\bf  r}_1,{\bf  r}_1)] + \Pi({\bf  r}_1,{\bf  r}_2),
\end{equation}
and for the correlation of amplitudes of two different eigenfunctions
\begin{equation} \label{fin2}
V^2\langle \vert \psi_k({\bf  r}_1) \psi_l({\bf  r}_2) \vert^2
\rangle_{E, \omega} -1  =  k_d(r) 
 \Pi({\bf  r}_1,{\bf  r}_1), \ k \ne l 
\end{equation}
In particular, for ${\bf  r}_1
= {\bf  r}_2$ we have 
\begin{equation} \label{fin3}
V^2 \langle \vert \psi_k ({\bf  r}) \psi_l {\bf  r}) \vert^2
\rangle_{E, \omega} - 1 = \delta_{kl} + (1 + \delta_{kl})
\Pi({\bf  r},{\bf  r}). 
\end{equation}
Note that the result (\ref{fin1}) for
${\bf  r}_1 = {\bf  r}_2$ is the inverse participation ratio
calculated in Ref. \cite{MF}; on the other hand, neglecting
the terms with the diffusion propagator (i.e. making the zero-mode
approximation), we reproduce the result of Refs.\cite{Berry,Prig,Srednicki}.  

Eqs. (\ref{fin2}), (\ref{fin3})  show
that the correlations between different eigenfunctions are relatively
small in the weak disorder regime. Indeed, they are proportional to
the small parameter $\Pi({\bf r},{\bf r})$, which is equal in the
case of 2D  geometry to ($L$ is the size of the system)
\begin{equation} 
\label{pi}
\Pi({\bf  r}, {\bf  r}) = 
 (\pi g)^{-1} \ln L/l,\qquad 2D,
\end{equation}
with $g=2\pi\nu D$. For a quasi-1D wire or strip of the length $L$,
\begin{equation}
\Pi(r,r)={2\over g}\left[{1\over 6}+ B_2\left({r\over L}\right)\right]\
,\qquad 0\le r\le L\ ,
\label{1d}
\end{equation}
where $g=2\pi\nu D/L$, and
$B_2(x)=x^2-x+1/6$ is the Bernoulli polynomial.\footnote{
In the 3D geometry,  the sum over the momenta $q$ in
Eq.(\ref{pi1}) determining $\Pi({\bf r},{\bf r})$ diverges at large
$q$ and is determined by the upper cut-off, $q\sim 1/l$, yielding
$\Pi({\bf r},{\bf r})\sim g^{-1}L/l$. This reflects the fact that in
3D geometry the truly local (${\bf r_1}={\bf r_2}$) correlations may
not be given correctly by the diffusion approximation and can depend
on microscopic structure of the random potential. For this reason, we
do not consider local correlations in 3D geometry here. Note, however,
that this concerns the global geometry of the sample; locally the
system can be either of 2D or 3D nature, which determines the form of
the function $k_d(r)$  (e.g, a wire is locally 3D, but has a quasi-1D
geometry).} 
The correlations are enhanced by disorder; when the system approaches
the strong localization regime, the
relative magnitude of correlations, $\Pi({\bf  r}, {\bf  r})$
ceases to be small. The correlations near the Anderson localization
transition will be discussed in the next section of the paper.

Another correlation function, generally used for the calculation of
the linear response of the system,
\begin{eqnarray} 
\label{gamma}
&& \gamma({\bf  r}_1, {\bf  r}_2, E, \omega) =
\left\langle \psi^*_k({\bf r}_1) \psi_l({\bf r}_1) 
\psi_k({\bf r}_2) \psi^*_l({\bf  r}_2) \right\rangle_{E,
\omega} \nonumber \\
&& \qquad\equiv  \Delta^2 R_2^{-1}(\omega)
\left\langle \sum_{k\ne l}  
\psi^*_k({\bf r}_1) \psi_l({\bf r}_1) 
\psi_k({\bf r}_2) \psi^*_l({\bf  r}_2)
\delta (E - \epsilon_k) \delta(E + \omega
- \epsilon_l) \right\rangle
\end{eqnarray}
can be calculated in a similar way; the result reads
\begin{equation} \label{states}
V^2 \langle \psi^*_k({\bf r}_1) \psi_l({\bf r}_1) 
\psi_k({\bf r}_2) \psi^*_l({\bf  r}_2) \rangle_{E, \omega}
 = k_d(r) + \Pi({\bf 
r}_1,{\bf  r}_2),\ \ \ k \ne l. 
\end{equation}

As is seen from Eqs. (\ref{fin1}), (\ref{fin2}), (\ref{states}),
 in the $1/g$ order the correlation functions 
$\alpha({\bf  r}_1, {\bf  r}_2, E)$
and $\gamma({\bf  r}_1, {\bf  r}_2, E, \omega)$
survive for the large separation between the points, $r \gg l$, 
while $\sigma({\bf  r}_1, {\bf  r}_2, E, \omega)$ decays
exponentially for the distances larger than the mean free path
$l$. This is, however, an artifact of the $g^{-1}$ approximation, and
the investigation of the corresponding tails requires the
extension of the above calculation to the terms proportional to
$g^{-2}$. We find that the correlator $\langle Q_{bb}^{11}({\bf
 r_1})Q_{bb}^{22}({\bf r_2})\rangle_F$ gets the following correction: 
\begin{eqnarray} \label{correct}
 \delta \langle Q_{bb}^{11}({\bf r_1})Q_{bb}^{22}({\bf r_2}) \rangle_F
&=& -f_1 + 2f_4 +\exp(2i\pi s) f_3 
 -2i {\exp(2i\pi s)\over \pi s} (f_2 - f_3) \nonumber \\
&-& {\exp(2i\pi s)-1\over 2(\pi s)^{2}} (f_1 - 4f_2 + 3f_3 - 2f_4). 
\end{eqnarray}
Here we defined the functions
\begin{eqnarray} \label{fun1}
f_1({\bf  r}_1, {\bf  r}_2) & = & \Pi^2 ({\bf  r}_1,
{\bf  r}_2), \nonumber \\
f_2({\bf  r}_1, {\bf  r}_2) & = & (2V)^{-1} \int d{\bf  r}
\left[\Pi^2 ({\bf  r}, {\bf  r}_1) + \Pi^2 ({\bf  r},
{\bf  r}_2) \right], \nonumber\\
f_3 & = & V^{-2} \int d{\bf  r}
d{\bf  r}' \Pi^2 ({\bf  r}, {\bf  r}'), \nonumber \\
f_4({\bf  r}_1, {\bf  r}_2) & = & V^{-1} \int d{\bf  r}
\Pi ({\bf  r}, {\bf  r}_1) \Pi ({\bf  r},
{\bf  r}_2). 
\end{eqnarray}
Consequently, we obtain the following results for the correlations of
different ($k\ne l$) eigenfunctions at $r>l$:
\begin{eqnarray} 
&&V^2\langle \vert \psi_k({\bf  r}_1) \psi_l({\bf  r}_2) \vert^2
\rangle_{E, \omega} -1   =  
{1\over 2}(f_1-f_3-2f_4) \nonumber\\
&&+2(f_2-f_3)\left({\sin^2 \pi s\over (\pi s)^2}-
{\sin 2\pi s\over 2\pi s}\right)
\left (1 - { \sin^2 \pi s\over (\pi s)^2}\right)^{-1}.
\label{corr2}
\end{eqnarray}
As it should be expected, the double integral over the both coordinates of 
this correlation function is equal to zero. This property is just the
normalization condition and should hold in arbitrary order of
expansion in $g^{-1}$.  

The quantities $f_2$, $f_3$, and $f_4$ are proportional to $g^{-2}$,
with some (geometry-dependent) prefactors of order unity. 
On the other hand, $f_1$ in $2D$ and $3D$ geometry depends
essentially on the distance $r=|{\bf r}_1-{\bf r}_2|$. In
particular, for $l\ll r\ll L$ we find
\begin{eqnarray*}
& & f_1({\bf r}_1,{\bf r}_2)=\Pi^2({\bf r}_1,{\bf r}_2)\approx
\left\{
\begin{array}{ll}
\displaystyle{{1\over (\pi g)^2}\ln^2{L\over r}\ }, & \qquad 2D\\
\displaystyle{{1\over (4\pi^2\nu Dr)^2}}        \ , & \qquad 3D
\end{array}
\right.
\end{eqnarray*}
Thus, for $l<r\ll L$, the contributions proportional to $f_1$ dominate
in Eq.(\ref{corr2}), and we get
\begin{equation}
V^2\langle \vert \psi_k({\bf  r}_1) \psi_l({\bf  r}_2) \vert^2
\rangle_{E, \omega} -1 = {1\over
2}\Pi^2({\bf r}_1,{\bf r}_2)\ ,\ \ k \ne l. 
\label{adm3}
\end{equation}
On the other hand, for the case of quasi-1D geometry (as well as in 2D
and 3D for $r\sim L$), all quantities $f_1$, $f_2$, $f_3$, and $f_4$
are of order of $1/g^2$. Thus, the correlator 
$\sigma({\bf r_1},{\bf r_2},E,\omega)$ acquires a
non-trivial (oscillatory) frequency dependence on a scale $\omega\sim\Delta$
described by the second term in the r.h.s. of Eq.(\ref{corr2}). In
particular, in the quasi-1D case the function $f_2-f_3$ determining
the spatial dependence of this term has the form
\begin{equation}
f_2-f_3=-{2\over 3g^2}\left[B_4\left({r_1\over L}\right)+
B_4\left({r_2\over L}\right)\right]\ ,
\label{f2f3}
\end{equation}
where $B_4(x)=x^4-2x^3+x^2-1/30$. 

Let us remind the reader that the above derivation is valid for $\omega \ll
E_c$. In order to obtain the results in the range $\omega \ge E_c$ one
can calculate the sigma-model correlation functions entering
Eqs. (\ref{e5}) by means of the perturbation
theory \refnote{\cite{AShk}}. We find then
\begin{eqnarray} \label{pert}
V^2\langle \vert \psi_k({\bf  r}_1) \psi_l({\bf  r}_2) \vert^2
\rangle_{E, \omega} &=& 1 + \mbox{Re} \left\{ k_d(r)
\Pi_{\omega}({\bf r}_1, 
{\bf r}_2) \right. \nonumber \\
 &+&\left.  \frac{1}{2} \left[ \Pi^2_{\omega} ({\bf r}_1,{\bf r}_2)
- \frac{1}{V^2} \int d{\bf r} d{\bf r}' \Pi^2_{\omega}
({\bf r},{\bf r}') \right] \right\}, \\
V^2 \langle \psi^*_k({\bf r}_1) \psi_l({\bf r}_1) 
\psi_k({\bf r}_2) \psi^*_l({\bf  r}_2) \rangle_{E, \omega} 
&=& k_d(r) + \mbox{Re} 
\Pi_{\omega} ({\bf r}_1,{\bf r}_2), \nonumber
\end{eqnarray}
where $\Pi_{\omega} ({\bf r}_1,{\bf r}_2)$ is the finite-frequency
diffusion propagator
\begin{equation} \label{Ask}
\Pi_{\omega} ({\bf r}_1,{\bf r}_2) = (\pi\nu)^{-1} \sum_{{\bf q}}
\frac{ \phi_q ({\bf r}_1) \phi_q ({\bf r}_2)}{Dq^2 - i\omega}, 
\end{equation}
and the summation in Eq. (\ref{Ask}) now includes ${\bf q} = 0$.
As was mentioned, the perturbation theory should give correctly the
non-oscillatory (in $\omega$) part of the correlation functions at
$\omega\gg\Delta$. Indeed, it can be checked that Eqs.(\ref{pert})
match the supersymmetric $\sigma$-model results in this
regime. Furthermore, in the $1/g$ order [which means keeping only linear
in $\Pi_\omega$ terms in (\ref{pert}) and neglecting $-i\omega$ in
denominator of Eq.(\ref{Ask})] Eqs.(\ref{pert}), (\ref{Ask}) reproduce
the exact results (\ref{fin1}), (\ref{states}) even at small
frequencies $\omega\sim\Delta$. We stress however that the perturbative 
calculation is not justified in this region and only the supersymmetry
method provides a rigorous derivation of these results. 

As was mentioned, the case of broken time reversal symmetry (unitary
ensemble) was considered above. Generalization to a system with
unbroken time 
reversal symmetry is straightforward \refnote{\cite{BM1}}; in
$1/g$-order Eqs.(\ref{fin1}), (\ref{fin2}), and (\ref{states}) are
modified as follows:
\begin{equation} \label{fin1o}
V^2\langle \vert \psi_k({\bf r_1}) \psi_k({\bf r_2}) \vert^2
\rangle_{E} 
= \left[ 1 + 2k_d (r) \right] \left[ 1 + 2 \Pi_D ({\bf r_1},
{\bf r_2}) \right],
\end{equation}
\begin{equation} \label{fin2o}
V^2\langle \vert \psi_k({\bf r_1}) \psi_l({\bf r_2}) \vert^2
\rangle_{E, \omega} - 1  =  2 k_d(r) \Pi_D ({\bf r_1},
{\bf r_2}) . 
\end{equation}
\begin{equation} \label{stateso}
 V^2 \langle \psi^*_k({\bf r_1}) \psi_l({\bf r_1}) 
\psi_k({\bf r_2}) \psi^*_l({\bf r_2}) \rangle_{E, \omega}
 = k_d(r) + \left[ 1 + k_d(r) \right]
\Pi_D({\bf r_1},{\bf r_2}),\ \ \ k \ne l.  
\end{equation}

Using the supersymmetry method, one can calculate also higher order
correlation functions of eigenfunction amplitudes. In particular, the
correlation function $\langle|\psi_k^4({\bf r_1})||\psi_k^4({\bf
r_2})|\rangle_E$ determines fluctuations of the inverse participation
ratio (IPR) $I_2=\int d{\bf r}|\psi^4({\bf r})|$. Details of the
corresponding calculation can be found in Ref.\cite{MF}; the result
for the relative variance of IPR, $\delta(I_2)=\mbox{var}(I_2)/\langle
I_2\rangle^2$ being
\begin{equation}
\delta(I_2)={8\over\beta^2}f_3={32 a_d\over\beta^2 g^2},
\label{varipr}
\end{equation}
with a numerical coefficient $a_d$ depending on the sample
dimensionality $d$ and equal to $a_1=1/90$, $a_2\approx 0.0266$ and
$a_3\approx 0.0527$ for quasi-1D, 2D, and 3D geometry
respectively. The fluctuations (\ref{varipr}) have the same relative
magnitude as the famous universal conductance fluctuations. Note also
that extrapolating Eq.(\ref{varipr}) to the Anderson transition point,
where $g\sim 1$, we find $\delta(I_2)\sim 1$, so that the magnitude of
IPR fluctuations is of the order of its mean value (which is, in turn,
much larger than in the metallic regime; see the next section).

Similarly, correlations of eigenfunction amplitudes determine
fluctuations of matrix elements of an operator of some (say, Coulomb) 
interaction computed on
eigenfunctions $\psi_k$ of the one-particle Hamiltonian in a random
potential. Such a problem naturally arises, when one wishes to study
the effect of interaction onto statistical properties of excitations in
a mesoscopic sample (see below).

Finally, the above consideration can be generalized to a ballistic chaotic
system, by applying a recently developed ballistic generalization of
the $\sigma$-model \refnote{\cite{MK,AASA}}. The results are then
expressed in terms of the (averaged over the direction of velocity)
kernel $g({\bf r_1}, {\bf n_1}; {\bf r_2}, {\bf n_2})$ of the
Liouville operator $\hat{K}=v_F{\bf n}\nabla$ governing the classical
dynamics in the system,
\begin{eqnarray}
\label{ball}
& & \Pi ({\bf r_1}, {\bf r_2})
= \int d{\bf n_1} d{\bf n_2}\, g({\bf r_1}, {\bf n_1}; {\bf r_2},
{\bf n_2}); \nonumber\\
& & \hat K g({\bf r_1}, {\bf n_1}; {\bf r_2}, {\bf n_2}) =
 \left( \pi \nu \right)^{-1} \left[ \delta({\bf r_1} - {\bf r_2})
\delta({\bf n_1} - {\bf n_2}) - V^{-1} \right].
\end{eqnarray} 
Here ${\bf n}$ is a unit vector determining the direction of momentum,
and normalization $\int d{\bf n}=1$ is used. 
Equivalently,  the function $\Pi({\bf r_1},{\bf r_2})$ can be defined
as
\begin{equation}
\label{ball2}
\Pi({\bf r_1},{\bf r_2})=\int_0^\infty dt\int d{\bf n_1}
\,\tilde{g}({\bf r_1},{\bf n_1},t;{\bf r_2})\ ,
\end{equation}
where $\tilde{g}$ is determined by the evolution equation
\begin{equation}
\label{ball3}
\left({\partial\over\partial t}+v_F{\bf n_1 \nabla_1}\right)
\tilde{g}({\bf r_1},{\bf n_1},t;{\bf r_2})=0\ ,\qquad t>0
\end{equation}
with the boundary condition
\begin{equation}
\label{ball4}
\tilde{g}|_{t=0}=(\pi\nu)^{-1}[\delta({\bf r_1}-{\bf r_2})-V^{-1}].
\end{equation}
Eq.(\ref{ball}) is a
natural ``ballistic'' counterpart of Eq.(\ref{diff}). In particular,
generalization of Eqs.(\ref{fin1}), (\ref{fin1o}) 
for the correlations of an eigenfunction amplitudes in two different
points  onto the ballistic case reads \refnote{\cite{BMM2}}
\begin{equation}
\label{ball1}
\alpha({\bf r_1}, {\bf r_2}, E) = 1 + {2\over \beta} \Pi({\bf r_1},
{\bf r_2}).
\end{equation}
Note that the ballistic $\sigma$-model approach is of semiclassical
nature and thus valid on distances much larger than the wave length
$\lambda_F$. For this reason, Eqs.(\ref{ball1}) represent the smoothed
correlation function, which is not valid for $|{\bf r_1}-{\bf r_2}|\le
\lambda_F$. Indeed, the contribution to $\Pi({\bf r_1},{\bf r_2})$
from the straight line motion from ${\bf r_2}$ to $\bf{r_1}$ can be
easily evaluated, yielding e.g. in the 2D case 
 $\Pi^{(0)}({\bf r_1},{\bf r_2})=1/(\pi p_F |\bf r_1-\bf r_2|)$. This
is nothing else but the smoothed version of the function 
$k_d(|{\bf r_1}-{\bf r_2}|)=J_0^2(p_F|{\bf r_1}-{\bf r_2}|)$ giving the
leading contribution to the short-scale correlations. A formula
for the variance of matrix elements closely related to
Eq.(\ref{ball1})
was obtained in the semiclassical approach in Ref.\cite{Eckhardt}.
In a very recent paper \refnote{\cite{Sred}} a similar
generalization of the Berry formula for $\langle\psi_k^*({\bf
r_1})\psi_k({\bf r_2})\rangle$ was proposed.

Eq.(\ref{ball1}) shows that correlations in eigenfunction amplitudes
in remote points are determined by the classical dynamics in the
system. It is closely related to the phenomenon of scarring of
eigenfunctions by the classical orbits \refnote{\cite{Heller,AF}}.
Indeed, if $\bf {r_1}$ and ${\bf r_2}$ belong to a short periodic
orbit, the function $\Pi({\bf r_1},{\bf r_2})$ is positive, so that
the amplitudes $|\psi_k({\bf r_1})|^2$ and $|\psi_k({\bf r_2})|^2$
are positively correlated. This is a reflection of the ``scars''
associated with this periodic orbits and a quantitative
characterization of their strength in the coordinate space. 
Note that this effect gets smaller
with increasing energy $E$ of eigenfunctions. Indeed, for a strongly
chaotic system and for $|{\bf r_1}-{\bf r_2}|\sim L$ ($L$ being the
system size), we have in the 2D case
$\Pi({\bf r_1},{\bf r_2})\sim\lambda_F/L$,
so that the magnitude of correlations decreases as $E^{-1/2}$. 
The function $\Pi({\bf r_1},{\bf r_2})$ was explicitly calculated 
in Ref.\cite{BMM2}
for a circular billiard with  diffusive surface scattering.

\section{STRONG CORRELATIONS OF EIGENFUNCTIONS AND LEVEL REPULSION
AT THE ANDERSON LOCALIZATION TRANSITION}

In the preceding section, we considered the metallic regime, where a 
typical wavefunction $\psi_i({\bf r})$ is extended and covers
uniformly all the sample volume. When system approaches the point of
Anderson transition $E_c$, these extended eigenfunctions become less and
less homogeneous in space showing regions with larger and smaller
amplitudes and eventually forming a multifractal structure in the
vicinity of $E_c$. To characterize the degree of
non-homogeneity quantitatively, it is convenient to use the inverse
participation ratio (IPR)
$ \langle I_2(E)\rangle=\int d{\bf r}\alpha({\bf r},{\bf r},E)$ 
For extended states this quantity is
inversely proportional to the system volume: $\langle
I_2(E)\rangle=A(E) 
L^{-d}$, with $L$ and $d$ standing for the system size and spatial
dimension, respectively. The coefficient $A$ in this relation
measures a fraction of the system volume where eigenfunction is
appreciably non-zero. In the regime of a good conductor $A\approx
1+2/\beta\sim 1$,
whereas close to the mobility edge $E=E_c$ it becomes large
and diverges like
$A(E)\propto |E-E_c|^{-\mu_2}$, with a critical index $\mu_2>0$ 
\refnote{\cite{Weg}}. This means that eigenfunctions become more and
more sparse, when the  system approaches the critical point of the
Anderson transition. Just at the mobility edge eigenfunctions
occupy a vanishing fraction of the system volume and IPR scales like
$\langle I_2(E)\rangle\propto L^{-d+\eta}$, with $\eta>0$. 
Such a behavior reflects
multifractal \refnote{\cite{CdC,Chalk,Huck}} structure of critical
eigenstates. At last, in the insulating phase any eigenstate
is localized in a domain of finite extension $\xi$ and IPR remains
finite in the limit of infinite system size $L\to \infty$.

This transparent picture serves as a basis for qualitative
understanding of spectral properties of disordered conductors.
Indeed, as long as eigenstates are well extended and cover the whole
sample, they overlap substantially and corresponding energy levels
repel each other in the same way as in RMT. As a result, the Wigner-Dyson
(WD) statistics describes 
well energy levels in a good metal \refnote{\cite{Efetov,AShk,KM}}.
In contrast, in the insulating phase different eigenfunctions
corresponding to levels close in energy are localized far apart from
one another and their overlap is negligible. This is the reason for
absence of correlations of energy levels in this regime -- the
so-called Poisson statistics.

However, a naive extrapolation of this argument to the vicinity of
the transition point would lead to a wrong  conclusion. Indeed, one might
expect that sparse (multifractal in the
critical point) eigenstates fail to overlap, that would result in  
essential weakening of level correlations close to the mobility edge and
vanishing level repulsion at $E=E_c$. However, numerical
simulations show \refnote{\cite{AS2,Shk,Eva,Braun,Zhar}} that even
at the mobility edge levels
repel each other strongly, though the whole statistics is
 different from the WD one. The purpose of this
section is to explain how this apparent contradiction is resolved.
We will  show that critical eigenstates for
nearby levels are so strongly correlated that they
 overlap well in spite of their sparse structure.

To calculate the overlap function $\sigma({\bf r},{\bf r},E,\omega)$
in the critical regime \refnote{\cite{FM2}}, 
we will exploit an exactly solvable model of
the Anderson transition -- so-called sparse random matrix (SRM)
model \refnote{\cite{sp1}}. This model is essentially equivalent to a
tight-binding model, which is locally of tree-like stricture (i.e has
no small-scale loops), with matrix size $N$ playing a role of the
system volume (number of sites of the tight-binding model). The SRM
model can be used to construct an 
effective mean-field theory of Anderson 
localization \refnote{\cite{FM-ema}} corresponding to the limit $d=\infty$.
The inverse participation ratio
$\langle I_2(E)\rangle=N \alpha({r}, {r},E)$ is
proportional to $1/N$ in the delocalized phase, as expected: 
$\langle I_2(E)\rangle=A(E)/N$.
The coefficient $A(E)$ diverges close to the
transition point, $|E-E_c|\ll E_c$, like $A(E)\propto
\exp{\left(\mbox{const} |E-E_c|^{-1/2}\right)}$ \refnote{\cite{FM-ema}}
which differs from the power-law behavior expected for conventional
 $d-$dimensional systems.
The origin of such a critical dependence was explained
in \refnote{\cite{FM-ema,MF-94}} and stems from the fact that $A(E)$ is
determined essentially by the  "correlation
 volume" $V(\xi)$ (i.e. number of sites at a
distance smaller than correlation length $\xi$), which is
exponentially large,
 $V(\xi)\propto\exp(\mbox{const}\,\xi)$, for tree-like structures,  
whereas
$V(\xi)\propto\xi^{d}$ for a $d-$dimensional lattice. Having this
difference in mind, one can translate  the results obtained
in the framework of $d=\infty$ models to their finite-dimensional
counterparts \refnote{\cite{MF-94}}.

The calculation of the inverse participation ration $N\alpha(
r, r,E)$ can be extended onto the overlap correlation function
$\sigma(r,r,E,\omega)$. Evaluating the r.h.s. of Eq.(\ref{e4}), we
find the following relation \refnote{\cite{FM2}}:
\begin{equation}
\sigma(r,r,E,\omega)=\frac{\beta}{\beta+2}\alpha(r,r,E)\ .
\label{9a}
\end{equation}
This relation is valid 
{\it everywhere} in the phase of extended eigenstates, up to
the mobility edge $E=E_c$, provided the number of sites (the system  
volume)
exceeds the correlation volume. In particular, it is valid
in the critical region $|E-E_c|\ll E_c$, where a typical eigenfunction
is very sparse and $\alpha(r,r,E)$ grows like
$\exp{\left(\mbox{const} |E-E_c|^{-1/2}\right)}$.

Eq.(\ref{9a}) implies the following structure of eigenfunctions within
an energy interval
$\delta E=\omega<A^{-1}(E)$.
Each eigenstate can be represented as a product
$\Psi_i(r)=\psi_i(r)\Phi_E(r)$.
Here the function $\Phi_E(r)$ is an eigenfunction envelope of "bumps and dips"
   It is the  same for
all eigenstates around  energy E, reflects underlying gross  
(multifractal)
spatial structure and governs the divergence of the inverse
participation ratio (i.e. of the factor $A(E)$)
at the critical point. In contrast, $\psi_i(r)$ is Gaussian white-noise
component fluctuating in space on the scale of lattice constant.
It fills in the envelope function $\Phi_E(r)$ in an individual way for
each eigenfunction, but is not critical, i.e. is not sensitive to  
the vicinity
of the Anderson transition. These Gaussian fluctuations  are
responsible for the factor $\beta/(\beta+2)$ (which is the same as in
the corresponding Gaussian Ensemble) in Eq.(\ref{9a}).

As was already mentioned, this picture is valid in the energy window
$\delta E\sim A^{-1}(E)$ around the energy $E$; the number of
levels in this window being large as $\delta E/\Delta\sim
NA^{-1}(E)\gg 1$ in the thermodynamic limit $N\to\infty$. These states
form a kind of Gaussian Ensemble on a spatially non-uniform
(multifractal for $E\to E_c$) background $\Phi_E(r)$. Since the
eigenfunction correlations are described by the formula (\ref{9a}),
which has exactly the same form as in the Gaussian Ensemble, it is not
surprising that the level statistics has the WD form
everywhere in the extended phase \refnote{\cite{sp1}}.

We believe on physical grounds that the same picture should hold for a
conventional $d$-dimensional conductor. First of all, the general
mechanism of the transition is the same in $d<\infty$ and $d=\infty$
models. Furthermore, the sparsity (multifractality) of eigenstates
near the transition point takes its extreme form for $d=\infty$ models
\refnote{\cite{MF-94}}, so that since the strong correlations (\ref{9a}) take
place at $d=\infty$ it would be very surprising if they do not hold
at finite $d$ as well. Finally, Eq.(\ref{9a}) is supported by the
results of the calculations in the weak localization regime. Keeping
only the leading (in $d\ge 2$) terms proportional to the powers of
$\Pi({\bf r},{\bf r})$ and considering the unitary ensemble for
definiteness, 
we have up to the two-loop order (see Eq.(\ref{corr2}) and
Ref.\cite{MF}) 
\begin{equation}
\label{wl}
\sigma({\bf r},{\bf r},E,\omega)=
V^{-2}\left[1+\Pi({\bf r},{\bf r})+{1\over 2}\Pi^2({\bf r},{\bf r})
+\ldots\right]={1\over 2}\alpha({\bf r},{\bf r}, E)\ ,
\end{equation}
in full agreement with Eq.(\ref{9a}).

Replacing $A(E)$ by the $d$-dimensional correlation volume
$\sim\xi^d$, we conclude that for $E$ close to $E_c$  Eq.(\ref{9a})
should be valid for $\omega<\Delta_\xi$, where $\Delta_\xi\propto
1/\xi^d$ is the level spacing in the correlation volume. For larger
$\omega$, $\sigma({\bf r},E,\omega)$ is expected to decrease as
$\omega^{-\eta/d}$ according to the scaling arguments
\refnote{\cite{Chalk,Huck,Pracz}}, so that we find
$\sigma({\bf r},E,\omega)/\alpha({\bf r},E)\sim
(\omega/\Delta_\xi)^{-\eta/d}$, up to a numerical coefficient of order
of unity. Again, for any value of the energy $E$ in the delocalized
phase, taking the system size $L$ large enough, $L\gg\xi$, we
have a large number of levels
${\delta E/\Delta}\sim{\Delta_\xi/\Delta} \propto (L/\xi)^d$
in the energy window $\delta E$ where Eq.(\ref{9a}) holds, so that
the level correlation will be of the WD form.

Finally, let us consider what happens when we go from the critical
regime ($\xi$ large, but $L\gg\xi$) to the critical point ($\xi\gg
L$). For this purpose, let us keep the system size $L$ fixed and
change the energy toward $E_c$, so that $\xi$ increases. When $\xi$
gets comparable to the system size, $\xi\sim L$, we have
$\Delta_\xi\sim\Delta$. This is the border of applicability of the
above consideration. Correspondingly, we find
\begin{equation}
\label{10}
\sigma({\bf r},{\bf r},E,\omega)/\alpha({\bf r},{\bf r},E)\sim 1\
 ,\qquad \omega<\Delta 
 \end{equation}
and $\sigma({\bf r},{\bf r},E,\omega)/\alpha({\bf r},{\bf r},E)\sim
 (\omega/\Delta)^{-\eta/d}$ for $\omega>\Delta$. When $E$ approaches
 further $E_c$, the correlation length $\xi\gg L$ gets irrelevant, so
 that these results will hold in the critical point ($\xi=\infty$). Of
 course, Eq.(\ref{10}) is not sufficient to ensure the WD statistics
in the critical point, since there is only of order of one level
 within its validity range $\delta E\sim\Delta$. Indeed, the
numerical simulations show that the level
 statistics on the mobility edge is different from the WD one
\refnote{\cite{AS2,Shk,Eva,Braun,Zhar}}.

However, Eq.(\ref{10}) allows us to make an important conclusion
concerning the behavior of $R_2(\omega)$ at small $\omega<\Delta$, or,
which is essentially the same, the behavior of the nearest neighbor
spacing distribution $P(s)$, $s=\omega/\Delta$, at $s<1$. For this
purpose, it is enough to consider only two neighboring levels. Let
their energy difference be $\omega_0\sim\Delta$. Let us now perturb
the system by a random potential $V({\bf r})$ with
$\langle V({\bf r})\rangle=0$,
$\langle
V({\bf r})V({\bf r'})\rangle=\Gamma\delta({\bf r}-{\bf r'})$.
For the two-level system it reduces to a $2\times 2$ matrix
$\{V_{ij}\},\ i,j=1,2$, with  elements $V_{ij}=\int d^d{\bf r}\,  
V({\bf r})
\Psi_i^*({\bf r})\Psi_j({\bf r})$. The crucial point is that the
variances of the diagonal and off-diagonal matrix elements are
according to Eq.(\ref{10}) equal
to each other up to a factor of order of unity:
\begin{equation}
\langle V_{11}^2\rangle/\langle|V_{12}^2|\rangle =
\sigma({\bf r},{\bf r},E,\omega)/\alpha({\bf r},{\bf r},E)\sim 1
\label{11}
\end{equation}
The distance between the perturbed levels is given by
$\omega=[(V_{11}-V_{22}+\omega_0)^2+|V_{12}|^2]^{1/2}$. Choosing the
amplitude of the  potential in such a way that the typical energy
shift $V_{11}\sim\Delta$ and using Eq.(\ref{11}), we find 
$\langle|V_{12}|^2\rangle\sim\Delta$.
As a result, the probability density for the level separation 
$\omega$ is for $\omega\ll\Delta$ of the form
$dP\sim(\omega/\Delta)^\beta d\omega/\Delta$, with some prefactor of
order of unity. We thus conclude that in the critical
point $P(s)\simeq c_\beta s^\beta$ for $s\ll 1$ with a coefficient
$c_\beta$ of order of unity, in agreement with the numerical 
findings~\refnote{\cite{Shk,Eva,Zhar}}.

\section{FLUCTUATIONS IN THE ADDITION SPECTRA OF QUANTUM DOTS}

In this section we discuss effects of the eigenfunction statistics on
the spectral properties of quantum dots. The electron levels of a
quantum dot can be resolved if the temperature $T$ is less than the
mean single-particle level spacing in a dot and can be studied in
transport experiments (see the recent review \refnote{\cite{dotrev}}
and references therein). In small dots containing
only few electrons these levels show a regular structure familiar from
the atomic physics \refnote{\cite{Tarucha}}. What we have in mind here
are however larger dots (containing in typical experiments from few
hundred to few thousand conducting electrons). These dots are either
disordered (diffusive) or, although being ballistic, are expected to
have chaotic dynamics due to their irregular shape. As a result, it is
natural to use a statistical description of the properties of energy
levels and eigenfunctions in such dots.

There are two types of the quantum dot spectra studied experimentally
via measuring their I--V characteristics:
(i) excitation spectrum, when excited levels are probed in a dot with
given number of electrons by increasing the source-drain voltage, and
(ii) addition spectrum, when electrons are added one by one by
changing the gate voltage. In the latter case, which will be the
subject of our consideration here, one finds narrow conductance peaks
separated the regions of (almost) zero current (Coulomb blockade). 
Statistics of the spacings between these peaks was studied in
a number of recent experiments \refnote{\cite{Sivan,Wharam,Markus}};
we will return to the experimental results below.

The simplest theoretical model which may be used to study distribution
of the spacings is as follows. One considers a dot as a fixed size
diffusive mesoscopic sample and assumes that changing a gate voltage
by an amount $\delta V_g$ simply reduces to a uniform change of the
potential inside the dot by a constant $\gamma\, \delta V_g$, 
with certain numerical coefficient $\gamma$ (``lever arm''). Such a
model was used for numerical simulations of the addition spectra in
Refs.\refnote{\cite{Sivan,Prus}}. Below we consider the statistics of
peak spacings within this model \refnote{\cite{BMM}}, and later return
to the approximations involved. We will neglect the spin degree of
freedom of electrons of first; inclusion of the spin will be also
discussed in the end of the section.

The distance between the two consecutive conductance peaks is given by
\begin{eqnarray}
S_N & = & (E_{N+2}-E_{N+1}) -  (E_{N+1}-E_{N}) \nonumber \\
    & = & \mu_{N+1}^{N+2} - \mu_N^{N+1}, \label{dot1}
\end{eqnarray}
where $E_N$ is the ground state of a sample with $N$ electrons. In the
second line of Eq.(\ref{dot1}) we rewrote $S_N$ in terms of the
Hartree-Fock single electron energy levels, with $\mu_i^j$ denoting
the energy of the state $\#j$ in the dot containing $i$ electrons. It is
convenient to decompose $S_N$ in the following way
\begin{eqnarray}
S_N & = & (\mu_{N+1}^{N+2}-\mu_{N}^{N+2}) +
(\mu_{N}^{N+2}-\mu_{N}^{N+1}) \nonumber \\
& \equiv & E_1 + E_2 
\label{dot2}
\end{eqnarray}
The quantity $E_2$ is the distance between the two levels of the same 
one-particle (Hartree-Fock) Hamiltonian $\hat{H}_N$ (describing a dot
with $N$ electrons) and is expected to obey RMT; in particular
$\langle E_2\rangle=\Delta$ and $\mbox{r.m.s.}(E_2)=a\Delta$ with a
numerical coefficient $a$ of order of unity [$a=0.52$ ($0.42$) for the
orthogonal (resp. unitary) ensemble]. On the other hand, $E_1$ is a
shift of the level $\#(N+2)$ due to the change of the Hamiltonian
$\hat{H}_N\to\hat{H}_{N+1}$ accompanying addition of the electron
$\#(N+1)$ to the system. It can be in turn decomposed into the following
three contributions \refnote{\cite{BMM}}
\begin{eqnarray}
E_1&=& e^2/C+\int d{\bf r}\left(|\psi_{N+1}^2({\bf r})|+
|\psi_{N+2}^2({\bf r})|\right)\delta U({\bf r}) \nonumber\\
&&  
+\int d{\bf r}d{\bf r'}|\psi_{N+1}^2({\bf r})||\psi_{N+2}^2({\bf r'})|
U_{\kappa}(|{\bf r}-{\bf r'}|) \nonumber \\
&=&E_1^{(0)}+E_1^{(1)}+E_1^{(2)}
\label{dot3}
\end{eqnarray}
Here $C$ is the dot capacitance, $\delta U({\bf r})$ is the change of
the self-consistent potential due to addition of one electron
(i.e. difference in the self-consistent potential in the dots 
with $N$ and $N+1$ electrons), and $U_\kappa({\bf r})$ is the 
screened Coulomb interaction (with the subscript $\kappa$ denoting the
inverse screening length).  In particular, in the experimentally
most relevant 2D case (which we will consider below) 
and assuming a circular form of the dot with
radius $R$, we have
\begin{equation}
\delta U({\bf r}) = - {e^2\over 2\kappa R}(R^2-r^2)^{-1/2},
\label{dot3a}
\end{equation}
while $U_\kappa$ is given in the Fourier space by
$\tilde{U}_\kappa({\bf q})=2\pi e^2/\epsilon(q+\kappa)$ with $\kappa=2\pi
e^2\nu/\epsilon$ and $\epsilon$ being the dielectric constant. 
The first term in Eq.(\ref{dot3}) (the
charging energy) determines the average value $\langle E_1\rangle$ and
thus the average peak spacing $\langle S_N\rangle$ (since $e^2/C\gg
\Delta$ for a large dot with $N\gg 1$). This is the only contribution
to $E_1$ which is kept by so-called constant interaction model,
which in addition neglects fluctuations of the capacitance $C$. 
Consequently, fluctuations of $S_N$ in the constant interaction model
are determined solely by fluctuations of the single-particle level
spacing $E_2$ and thus should be described by RMT:
$\mbox{r.m.s.}(S_N)=a\Delta$. 

The term $E_1$ in Eq.(\ref{dot2}) is however an additional source of
fluctuations and is thus responsible for the enhancement of
fluctuations in comparison with RMT. In principle, all three terms 
$E_1^{(0)}$, $E_1^{(1)}$, and $E_1^{(2)}$ in Eq.(\ref{dot3})
contribute to this enhancement. Fluctuations of the first one,
$E_1^{(0)}=e^2/C$ are due to the fact that the capacitance is slightly
different from its purely geometric value because of a finite value of
the screening length. The corresponding correction to $C$ can be
expressed in terms of the polarization operator $P({\bf r},{\bf r'})$
\refnote{\cite{BM1}}. The latter is a fluctuating quantity (because of
fluctuations of the eigenfunctions in the Fermi sea) and contains
a random part $P_r({\bf r},{\bf r'})$ leading to the following
expression for the random part of the charging energy:
\begin{equation}
(e^2/C)_r=2\int d{\bf r} d{\bf r'}\delta U({\bf r})P_r({\bf r},{\bf
r'}) \delta U({\bf r'}).
\label{dot4}
\end{equation}
Evaluating the fluctuations of the polarization operator
\refnote{\cite{BM1}}, we find \footnote{We use the obvious notations
for the variance and the root mean square deviations of a quantity
$X$: $\mbox{var}(X)=\langle X^2\rangle - \langle X\rangle^2$;
$\mbox{r.m.s.}(X)=[\mbox{var}(X)]^{1/2}$.}
\begin{eqnarray}
\mbox{var}(E_1^{(0)}) &=& {48\over\beta}\nu^2\ln g\left[{1\over V}\int
d{\bf r_1} d{\bf r_2}\delta U({\bf r_1})\Pi({\bf r_1},{\bf
r_2}) \delta U({\bf r_2})\right]^2 \nonumber \\
& \propto & {1\over \beta} \ln g \left({\Delta\over g}\right)^2
\label{dot5}
\end{eqnarray}

Now we consider fluctuations of the last term, $E_1^{(2)}$, in
Eq.(\ref{dot3}). Using Eqs.(\ref{fin1}), (\ref{fin1o}) 
for the correlations of
eigenfunction amplitudes in two remote points, the variance of
$E_1^{(2)}$ is found to be
\begin{eqnarray} 
\mbox{var}(E_1^{(2)}) & = & {4\over \beta^2 V^4}\int
d{\bf r_1}d{\bf r'_1}d{\bf r_2}d{\bf r'_2}U_\kappa(|{\bf r_1}-{\bf r'_1}|)
U_\kappa(|{\bf r_2}-{\bf r'_2}|)\Pi({\bf r_1},{\bf r_2}) \Pi({\bf
r'_1},{\bf r'_2}) \nonumber\\ 
&\approx &{4\Delta^2\over \beta^2 V^2}\int d{\bf r_1}d{\bf r_2}
\Pi^2({\bf r_1},{\bf r_2}) \nonumber \\
&\propto & {1\over \beta^2}\left({\Delta\over g}\right)^2.
\label{dot6}
\end{eqnarray}
Finally, fluctuations of the term $E_1^{(1)}$ can be also evaluated
with help of Eqs.(\ref{fin1}), (\ref{fin1o}), yielding
\begin{eqnarray}
\mbox{var}(E_1^{(1)})&= &{4\over \beta V^2}\int d{\bf r_1}d{\bf r_2} 
\delta U({\bf r_1})\Pi({\bf r_1},{\bf
r_2}) \delta U({\bf r_2})
\nonumber \\
&\propto &{1\over\beta}{\Delta^2\over g}.
\label{dot7}
\end{eqnarray}

It is seen that for $g\gg 1$
all the contributions Eqs.(\ref{dot5})--(\ref{dot7})
are parametrically small compared to the RMT fluctuations (which are
$\sim\Delta$). Fluctuations of the term $E_1^{(1)}$ related to the
change $\delta U({\bf r})$ of the self-consistent potential represent
parametrically leading contribution 
 to the enhancement of the  peak spacing fluctuations with respect to
RMT. 

Let us now discuss approximations made in the course of the above
derivation:
\begin{enumerate}
\item[i)] It was assumed that changing of the gate voltage results in a
spatially uniform change of the potential in the sample, that led us to
the expression Eq.(\ref{dot3a}) for the change of the self-consistent
potential $\delta U({\bf r})$ accompanying the addition of one
electron to the 
dot. This result would correspond to a gate located far enough from
the sample. In a more realistic situation, when the gate is relatively
narrow and located close to the sample, the potential change $\delta
U({\bf r})$ (as well as the additional electron density) will be
mainly located on the side of the dot facing the gate. Furthermore,
the total area of the dot is not fixed, so that 
the dot gets larger (and slightly deformed) with adding each electron
to it. These effects lead to some increase of the fluctuations of
$E_1^{(1)}$ in comparison with the model considered above. If
the size of the gate and its distance to the dot are of the same
order of magnitude as the dot size, then
$$\mbox{r.m.s.}(E_1^{(1)})\propto \Delta/\sqrt{\beta g},$$ as in
Eq.(\ref{dot7}), with a geometry dependent-numerical prefactor. The
upper bound for the magnitude of fluctuations of $E_1^{(1)}$ is given
by the opposite limiting case, when additional electron density
occupies an area $\sim\lambda_F\times\lambda_F$ near the gate, in
which case one finds
$$\mbox{r.m.s.}(E_1^{(1)})\sim\Delta/\sqrt{\beta}.$$

\item[ii)] The dot was supposed to be diffusive in the
calculation. For a ballistic dot one should replace $\Pi({\bf r},{\bf
r'})$ by $\Pi_B({\bf r},{\bf r'})$, as was explained above. This would
mean that the parameter $g$ is replaced by $\sim
N^{1/2}\sim L/\lambda_F$, where $N$ 
is the number of electrons in the dot and $L$ the characteristic
linear dimension. The numerical coefficient would
depend, however, on ``how strongly chaotic'' is the dot. Role of the
eigenfunctions fluctuations and correlations (``scars'') in
enhancement of the peak spacing fluctuations was studied 
in \refnote{\cite{Stopa}} via numerical
simulations of a dot with $N\approx 100$ electrons. 

\item[iii)] It was assumed that the dot energy and the measured gate
voltage are related through a constant (or smoothly varying)
coefficient $\gamma$. This ``lever arm'' $\gamma$ depends, however on
the dot-gate capacitance, which is also a fluctuating quantity. If the
gate size and the distance to the gate is of the same order as the
size of the dot, these fluctuations should be of the same order as
fluctuations of the dot self-capacitance given by Eq.(\ref{dot5}),
and thus lead to additional fluctuations which are parametrically
 small compared to $\Delta$. In a more general situation (thin gate
located close to the sample) an additional analysis along the lines of
Ref.\cite{BM1} is necessary.

\item[iv)] The calculation was done within the random phase
approximation, which assumes that $r_s\equiv {e^2/\epsilon v_F}\ll
1$, with $v_F$ being the Fermi velocity. 
In realistic dots however $r_s\sim 1$. Since this value is still
considerably lower than the Wigner crystallization threshold, the
calculations should be still valid, up to a numerical factor
$\alpha(r_s)$ [depending on $r_s$ only and such that $\alpha(r_s\ll
1)=1$]. 

\item[v)] We considered the model of spinless electrons up to now. 
Let us briefly discuss the role of the spin degree of freedom. 
Within the constant interaction model, it would lead to a bimodal
distribution \refnote{\cite{Prus}} of peak spacings 
\begin{equation}
{\cal P}(S_N)={1\over 2}\left[\delta(S_N-e^2/C)+
{1\over 2\Delta}P_{WD}\left({S_N-e^2/C\over
2\Delta}\right) \right],
\label{dot8}
\end{equation}
where $P_{WD}(s)$ is the Wigner-Dyson distribution and $\Delta$ denotes
the level spacing in the absence of spin degeneracy. 
The value of the
coefficient $a$ in the relation $\mbox{r.m.s.}(S_N)=a\Delta$ is then
increased (compared to the spinless case)  and is equal to 1.24 (1.16)
for the 
orthogonal (resp. unitary) ensemble. Taking into account fluctuations
of eigenfunctions (and thus of $E_1$) 
however modifies the form of the distribution
\refnote{\cite{BMM}}. The value of 
the term $E_1^{(2)}$ representing the interaction between two
electrons is larger in the case when $\psi_{N+2}$ and $\psi_{N+1}$
correspond to two spin-degenerate states (i.e. have the same spatial
dependence of the wave function), since
\begin{eqnarray}
\label{dot9}
&&\langle\int d{\bf r}d{\bf r'}|\psi_{i}^2({\bf r})||\psi_{i}^2({\bf r'})|
U_{\kappa}(|{\bf r}-{\bf r'}|)\rangle-
\langle\int d{\bf r}d{\bf r'}|\psi_{i}^2({\bf r})||\psi_{j}^2({\bf r'})|
U_{\kappa}(|{\bf r}-{\bf r'}|)\rangle \nonumber \\
&&\hspace{1cm}
={2\over\beta V^2}\int d{\bf r}d{\bf r'}k_d(|{\bf r}-{\bf r'}|)
U_{\kappa}(|{\bf r}-{\bf r'}|)\sim \Delta
\end{eqnarray}
for $r_s\sim 1$ (the coefficient depends on $r_s$, see
\refnote{\cite{BMM}}). Therefore, filling a state $\psi_{i\uparrow}$ pushes
up the level $\psi_{i\downarrow}$ (with respect to other eigenstates) 
by an amount of order of
$\Delta$. This removes a bimodal structure of the distribution of peak
spacings and slightly modifies the value of the coefficient $a$. 

\end{enumerate}

Basing on the above analysis, we can make the following general
statement. Imaging that we fix $r_s\sim 1$ (i.e. fix the electron
density and thus the Fermi wave length) and the system geometry, and then
start to increase the linear dimension $L$ of the system. Then, while
the average value of the peak spacing $S_N$ scales as $\langle
S_N\rangle\approx e^2/C\propto 1/L$, its fluctuations will scale
differently: $\mbox{r.m.s.}(S_N)\sim\Delta\propto 1/L^2$. This result
is not at all trivial, since in an analogous problem for classical
particles \refnote{\cite{class,Koulakov}}
the fluctuations are proportional to the mean value 
$\langle S_N\rangle$. The physical reason for  smaller
fluctuations in the quantum case is in the delocalized nature of the
electronic wave functions, which are spread roughly uniformly over the
system. 

The above prediction was confirmed by a recent experiment
\refnote{\cite{Markus}}, where a thorough study of the peak spacing
spacing statistics was carried out. It was found that the
low-temperature value of $\mbox{r.m.s.}(S_N)$, as well the typical
temperature scale for its change are approximately given by the mean
level spacing $\Delta$ (while in units of $E_c$ the magnitude of
fluctuations was very small, typically 2--4\%). 
We note also that in recent numerical
simulations \refnote{\cite{Stopa}} fluctuations of the addition
energies were found to be approximately $0.7\Delta$, in agreement
with our results. 

\section{ACKNOWLEDGMENTS}

It is a pleasure to thank my collaborators Ya.~M.~Blanter, Y.~V.~Fyodorov,
and B.~A.~Muzykantskii. Useful discussions with
O.~Agam, S.~Fishman, Y.~Gefen, V.~E.~Krav\-tsov, and C.~Marcus are
gratefully  acknowledged. I also thank S.~R.~Patel for a useful remark.
This research was supported by SFB195 der Deutshen Forschungsgemeinschaft.

\end{document}